\documentclass[english,aps,pre,reprint,showpacs,amsfonts,amsmath,amssymb,nofootinbib,a4paper]{revtex4-1}
\usepackage[english]{babel}
\usepackage{amsfonts}
\usepackage{amsmath}
\usepackage{amssymb}
\usepackage{graphicx}
\usepackage{epstopdf}
\usepackage{enumerate}
\usepackage{color}

\begin{document}

\def\arXiv#1#2#3#4{{#1} #2 #3 {\it Preprint} #4}
\def\Book#1#2#3#4#5{{#1}, {\it #3} (#4, #5, #2).}
\def\Bookwd#1#2#3#4#5{{#1} {\it #3} (#4, #5, #2)}
\def\Journal#1#2#3#4#5#6#7{#1, #4 \textbf{#5}, #6 (#2).}
\def\JournalE#1#2#3#4#5#6{#1, #4 \textbf{#5}, #6 (#2).}
\def\eref#1{(\ref{#1})}

\newcommand{\dd}{\mbox{d}}
\newcommand{\EE}{\mathbb{E}}
\newcommand{\NN}{\mathbb{N}}
\newcommand{\PP}{\mathbb{P}}
\newcommand{\RR}{\mathbb{R}}
\newcommand{\TT}{\mathbb{T}}
\newcommand{\ZZ}{\mathbb{Z}}
\newcommand{\uu}{\mathbf{1}}
\newcommand{\HH}{\mathcal{H}}

\title{
Classical-like behavior in quantum walks with inhomogeneous, time-dependent coin operators}
\author{Miquel Montero}
\email{miquel.montero@ub.edu}
\affiliation{Departament de F\'{\i}sica Fonamental, Universitat de Barcelona (UB), Mart\'{\i} i Franqu\`es 1, E-08028 Barcelona, Spain}
\altaffiliation[Current address: ]{Secci\'o de F\'{\i}sica Estad\'{\i}stica i Interdisciplin\`aria, Departament de F\'{\i}sica de la Mat\`eria Condensada, Universitat de Barcelona (UB), Mart\'{\i} i Franqu\`es 1, E-08028 Barcelona, Spain}
\date{\today}

\pacs{03.67.-a, 05.40.Fb}

\begin{abstract}
Although quantum walks exhibit peculiar properties that distinguish them from random walks, classical behavior can be recovered in the asymptotic limit  by destroying the coherence of the pure state associated to the quantum system. Here I show that this is not the only way: I introduce a quantum walk driven by an inhomogeneous, time-dependent coin operator, which mimics the statistical properties of a random walk at all time scales. The quantum particle undergoes unitary evolution and, in fact, the high correlation evidenced by the components of the wave function can be used to revert the outcome of an accidental measurement of its chirality. 
\end{abstract}
\maketitle

\section{Introduction}

Quantum walks (QWs)~\cite{VA12} were originally termed ``quantum random walks''~\cite{ADZ93,TM02,NK03,JK03} as they were thought as the quantum-mechanical version of the discrete random walk (RW) in one dimension: the Markov process in which a particle changes its position at each clock tick by jumping to one of the two nearest sites depending on the random outcome of a coin toss. This source of randomness could be seen as superfluous in the quantum world, where the location of a particle is a probabilistic magnitude, governed by its wave function. Therefore, in the design of these ``quantum random walks'', the coin toss was replaced by some (unitary) operator that affects the state of a quantum binary property of the system, e.g., the spin or the chirality, and the wave function is shifted according to the value of this qubit.

Consequently, beyond the intrinsic uncertainty of the quantum phenomena, ``quantum random walks'' are not random at all |and thus this term is now deprecated. The most prominent sign of this {\it deterministic\/} nature of QWs is the ballistic behavior they can show~\cite{ABNVW01}, the ability to connect any two sites after a lapse of time that is proportional to the distance between these sites, even if the walk is undirected. This fact comes in conflict with the diffusive nature of unbiased RWs which, to perform the same operation, need a time period that grows quadratically with the separation of the sites. This speed-up readily caught the attention of the scientific community, albeit there are other properties that distinguish QWs from RWs~\cite{CFG03}. In spite of those differences, QWs are indeed the quantum analogues of RWs, and therefore they may experience a change from ballistic to diffusive motion when the quantum coherence of the state is affected by multiple reasons~\cite{BCA03a,BCA03b,KT03,WD03,SCPGJS11}. In fact, it has been proven that, under mild conditions, the introduction of temporal or spatial fluctuations in the properties of the coin operator acting upon the QW leads to classical behavior in the asymptotic limit: the standard deviation of the position of the quantum walker grows with the square root of the elapsed time, and the corresponding rescaled distribution converges to a Gaussian~\cite{AVWW11,ACMSWW2}.

Soon after the birth of the very concept of quantum computers~\cite{RPF85}, i.e., computers whose operation cannot be understood without the laws of quantum mechanics~\cite{DAM96}, the first genuine quantum algorithms appeared~\cite{PS97,LKG97,FG98}, algorithms that were more efficient than their classic counterparts. And since many of those classical algorithms use RWs as building blocks, it is not surprising that the ballistic transport of QWs was seen as the key feature in the design of faster algorithms~\cite{SKW03,AMB10,MNRS11}. But QWs can play an even more important role in quantum computation, as they may be regarded as universal computational primitives~\cite{AC09,LCETK10}, i.e., they can be used to implement all the logic gates that a universal quantum computing machine needs to work~\cite{DD85}. 

While it is still an open issue whether universal quantum computers can efficiently simulate arbitrary physical systems~\cite{NC10}, in this paper I will give an affirmative answer to a related but much more limited question: Can a quantum walk be used to simulate the behavior of a classical system whose evolution follows a random walk? 

Specifically, I will look for a QW that shows {\it exactly\/} the same probabilistic properties of a RW at all time scales. This objective must not be reached as a result of the introduction of exogenous disturbances that can induce decoherence in the pure quantum state: The purpose is to obtain the classical distribution (the binomial distribution) by means of reversible unitary evolution at every time step. Therefore, this QW with {\it classical-like\/} attributes could replace the corresponding RW in the simulation of the dynamics of the desired classical system on a quantum computer. 

With this aim, I consider here a discrete-time QW on the line endowed with an inhomogeneous, time-dependent coin operator. Extensions of this kind were considered in the past: one can find in the literature examples of QWs driven by inhomogeneous, site-dependent coins~\cite{BFT08,SK10,KLS13,ZXT14,XQTS14}, time-dependent coins~\cite{RMM04,BNPRS06,AR09a,AR09b,MM14}, or history-dependent coins~\cite{FAJ04,SWH14}.

The paper is organized as follows. Section~\ref{Sec_Process} reviews the formalism used in the construction of the discrete-time quantum walk on the line with a time-dependent coin operator. Section~\ref{Sec_Unbiased} shows how one can devise a QW that behaves like an unbiased RW. I extend the framework to encompass general RWs in Sec.~\ref{Sec_Biased}. Section~\ref{Sec_Redundancy} explores the possibility of reverting the consequences of a measurement of the chirality, and quantifies the entanglement between the chirality and the position of the particle by means of von Neumann's entropy. When this correlation is destroyed, Sec.~\ref{Sec_Decoherence}, the walker becomes vulnerable to decoherence.
The paper ends with Sec.~\ref{Sec_Conclusion}, where conclusions are drawn.

\section{QW with an inhomogeneous, time-dependent coin operator}
\label{Sec_Process}
I begin the discussion by introducing the foundations of the inhomogeneous, time-dependent quantum walk on the line. I denote by $\HH^{\text p}$ the Hilbert space of discrete particle positions in one dimension, spanned by the basis $\left\{| n\rangle : n \in \ZZ\right\}$, and by $\HH^{\text c}$ the Hilbert space of coin states, spanned by the basis $\left\{|+\rangle, |-\rangle\right\}$. The discrete-time, discrete-space quantum walk on the Hilbert space $\HH\equiv\HH^{\text c}\otimes \HH^{\text p}$ is the result of the action of the evolution operator $\widehat{T}_t$, $\widehat{T}_t\equiv \widehat{S}\, \widehat{U}_t$, where the {\it coin\/} $\widehat{U}_t$ is an inhomogeneous, time-dependent, real-valued unitary operator:
\begin{eqnarray}
\widehat{U}_t
&\equiv& \sum_{n=-\infty}^{\infty}\big[\cos \theta^{ }_{n,t} |+\rangle  \langle +| + \sin \theta^{ }_{n,t} |+\rangle  \langle -| \nonumber \\
&+&\sin \theta^{ }_{n,t}  |-\rangle  \langle +| - \cos \theta^{ }_{n,t}  |-\rangle  \langle -|\big]\otimes |n\rangle  \langle n|,
\label{U_coin_gen}
\end{eqnarray}
with $0\leq \theta^{ }_{n,t}\leq \pi$, and $\widehat{S}$ is the shift operator that {\it moves\/} the walker depending on the respective coin state:
\begin{equation}
\widehat{S} |\pm\rangle\otimes| n\rangle = |\pm\rangle\otimes| {n\pm 1}\rangle.
\end{equation}
As the time increases in discrete steps, one chooses the time units so that the time variable $t$ is just an integer index, and the state of the system at a later time, $|\psi\rangle_{t+1}$, is recovered by applying $\widehat{T}_t$ to the present state $|\psi\rangle_{t}$:   
\begin{equation}
|\psi\rangle_{t+1} =\widehat{T}_t|\psi\rangle_{t}.
\label{evol_t}
\end{equation}
Equation~\eqref{evol_t} induces the following set of recursive equations:
\begin{eqnarray}
\psi^{ }_{+}(n+1,t+1)&=&\cos \theta^{ }_{n,t} \,\psi^{ }_{+}(n,t)+\sin \theta^{ }_{n,t} \,\psi^{ }_{-}(n,t),\nonumber\\\label{Rec_P}
\\
\psi^{ }_{-}(n-1,t+1)&=&\sin \theta^{ }_{n,t} \,\psi^{ }_{+}(n,t)-\cos \theta^{ }_{n,t} \,\psi^{ }_{-}(n,t),\nonumber\\
\label{Rec_M}
\end{eqnarray}
on the wave-function components, $\psi_{\pm}(n,t)$, the  projections of the state of the walker into the elements of the basis of the Hilbert space:
\begin{eqnarray}
\psi^{ }_{+}(n,t)&\equiv& \langle n|  \otimes  \langle +| \psi\rangle_t, \label{Def_Psi_P}\\
\psi^{ }_{-}(n,t)&\equiv& \langle n|  \otimes  \langle -| \psi\rangle_t. \label{Def_Psi_M} 
\end{eqnarray}

The evolution of the system is fully determined once $|\psi\rangle_{0}\equiv|\psi\rangle_{t=0}$ is set. Since the final aim is to simulate a RW, we must consider that the particle is initially located at the origin. When the coin operator is homogenous and time-independent, it is well known that the chirality of such localized state affects the ulterior behavior of  the system~\cite{TFMK03,MM15}. In our case, as we will see later on, this choice is not so delicate. Thus, for the sake of simplicity, I assume that there is no preferred direction in the chirality:
\begin{equation}
|\psi\rangle_{0} =\frac{1}{\sqrt{2}}\left(|+\rangle +   |-\rangle\right) \otimes | 0\rangle,
\label{Psi_Zero_Gen}
\end{equation}
that is $\psi_{\pm}(0,0)=1/\sqrt{2}$. Note that a real-valued state at time $t=0$ precludes the possibility of having a complex-valued wave function at a later time, cf. Eqs.~\eqref{Rec_P} and~\eqref{Rec_M}.

We want to connect the evolution of our quantum system with the statistical properties of a random walker. This connection must be done through the pairing of the probability mass function (PMF) of the two processes: Let us call $\rho(n,t)$ the likelihood of finding the particle in a particular position $n$ at a given time $t$. In the case of the quantum walker this probability depends on the wave-function components,
\begin{eqnarray}
\rho(n,t)&\equiv&\left|\psi^{ }_{+}(n,t)\right|^2+\left|\psi^{ }_{-}(n,t)\right|^2.
\end{eqnarray}
Our goal is to get that $\rho(n,t)$ equals the PMF of a random walk,
the binomial distribution:
\begin{eqnarray}
\rho(n,t)&=&\frac{t!}{\left(\frac{t+n}{2}\right)! \left(\frac{t-n}{2}\right)!}p^{\frac{t+n}{2}}\left(1-p\right)^{\frac{t-n}{2}},
\label{Rho_Gen}
\end{eqnarray}
for $n\in\left\{-t,-t+2,\cdots,t-2,t\right\}$. Function $\rho(n,t)$ determines the different moments of the stochastic process,
\begin{equation*}
\langle \widehat{X}^k_{ }  \rangle_t^{ } \equiv\sum_{n=-t}^{t} n^k \rho(n,t),
\end{equation*}
among which are worth to be highlighted the expectation value of the walker position, $\langle \widehat{X} \rangle_t^{ }$, and its uncertainty $\Delta X_t^{ }$, magnitudes that should amount to
\begin{eqnarray}
\langle \widehat{X} \rangle_t^{ } &=&(2 p-1) t,\\
\Delta X_t^{ }\equiv\sqrt{\langle \widehat{X}^2_{ } \rangle_t^{ }-\langle \widehat{X} \rangle^2_t} &=&2\sqrt{p(1- p) t},
\end{eqnarray}
if the classical expression~\eqref{Rho_Gen} is valid. 
Note that, since the probability is conserved, the change in the expectation value of the position at consecutive instants reads
\begin{eqnarray}
\langle \widehat{X} \rangle_{t+1}^{ }&=&\langle \widehat{X} \rangle_{t}^{ }
+\sum_{n=-t}^{t} J(n,t),
\label{Ehrenfest}
\end{eqnarray}
where $J(n,t)$,
\begin{eqnarray}
J(n,t)&\equiv& \cos 2\theta^{ }_{n,t}\left[\psi^{2}_{+}(n,t)-\psi^{2}_{-}(n,t)\right]\nonumber\\
&+&2\sin 2\theta^{ }_{n,t}\psi^{ }_{+}(n,t) \psi^{ }_{-}(n,t),
\label{J_def}
\end{eqnarray}
is the net flux of probability entering or leaving site $n$, and its explicit expression stems from Eqs.~\eqref{Rec_P} and~\eqref{Rec_M}. As we will see in the next section, Eqs.~\eqref{Ehrenfest} and~\eqref{J_def} pave the way for achieving our purpose.

\section{Unbiased walk}
\label{Sec_Unbiased}

\begin{figure}[htbp]
\includegraphics[width=0.9\columnwidth,keepaspectratio=true]{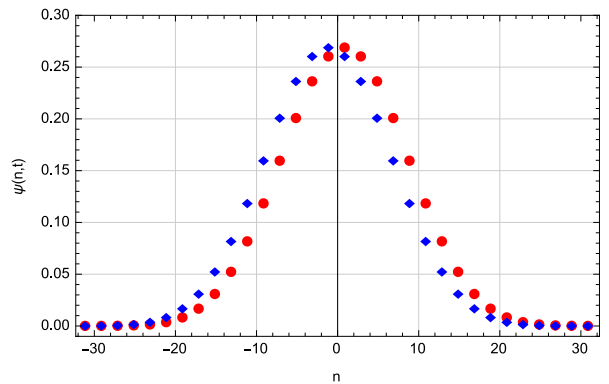}\
\caption{(Color online) 
The two components of the wave function at $t=31$. The red dots correspond to $\psi^{ }_{+}(n,t)$ whereas the blue diamonds mark the values of $\psi^{ }_{-}(n,t)$.} 
\label{Fig_Unbiased}
\end{figure}

Our task is therefore to deduce a functional form for $\cos \theta^{ }_{n,t}$ that can be accommodated in Eqs.~\eqref{Rec_P} and~\eqref{Rec_M} and ultimately lead to the desired PMF, Eq.~\eqref{Rho_Gen}. In order to grasp the appropriate procedure, I will consider the unbiased version of the RW in the first place, 
\begin{eqnarray}
\rho(n,t)=\frac{1}{2^t}\frac{t!}{\left(\frac{t+n}{2}\right)! \left(\frac{t-n}{2}\right)!},
\label{Rho_Sym}
\end{eqnarray}
for $n\in\left\{-t,-t+2,\cdots,t-2,t\right\}$. This results in a great simplification since in this case the expectation value of the position is null , $\langle \widehat{X} \rangle_t^{ }=0$, for any time value. This property is preserved by Eq.~\eqref{Ehrenfest} if $J(n,t)=0$, a sufficient condition. The absence of probability flux can be readily achieved, see Eq.~\eqref{J_def}, if
\begin{eqnarray}
\cos 2\theta^{ }_{n,t} &=& -\frac{2\psi^{ }_{+}(n,t) \psi^{ }_{-}(n,t)}{\rho(n,t)},\\
\sin 2\theta^{ }_{n,t} &=& \frac{\psi^{2}_{+}(n,t)-\psi^{2}_{-}(n,t)}{\rho(n,t)},\end{eqnarray}
that is,
\begin{eqnarray}
\cos \theta^{ }_{n,t} &=&  \frac{1}{\sqrt{2}}\frac{\psi^{ }_{+}(n,t)-\psi^{ }_{-}(n,t)}{\sqrt{\rho(n,t)}},\label{Cos_Sym}\\
\sin \theta^{ }_{n,t} &=&  \frac{1}{\sqrt{2}}\frac{\psi^{ }_{+}(n,t)+\psi^{ }_{-}(n,t)}{\sqrt{\rho(n,t)}}\label{Sin_Sym}.
\end{eqnarray}
It is easy to check that Eqs.~\eqref{Cos_Sym} and~\eqref{Sin_Sym} represent valid trigonometric expressions. Now, one can introduce these formulas in Eqs.~\eqref{Rec_P} and~\eqref{Rec_M} and obtain:
\begin{eqnarray}
\psi^{ }_{+}(n+1,t+1)=\psi^{ }_{-}(n-1,t+1)= \sqrt{\frac{\rho(n,t)}{2}},
\label{Redund_Sym}
\end{eqnarray}
leading to
\begin{eqnarray}
\psi^{ }_{+}(n,t)&=& \sqrt{\frac{(t-1)!}{2^t \left(\frac{t+n-2}{2}\right)! \left(\frac{t-n}{2}\right)!}}, \label{Sol_Sym_Psi_P}\\
\psi^{ }_{-}(n,t)&=&  \sqrt{\frac{(t-1)!}{2^t \left(\frac{t+n}{2}\right)! \left(\frac{t-n-2}{2}\right)!}}, \label{Sol_Sym_Psi_M} 
\end{eqnarray}
for $n\in\left\{-t+2,-t+4,\cdots,t-4,t-2\right\}$, and
\begin{eqnarray}
\psi^{ }_{+}(t,t)&=&\psi^{ }_{-}(-t,t)=\left(\frac{1}{2}\right)^{\frac{t}{2}},\label{Sol_Sym_End_a}\\
\psi^{ }_{+}(-t,t)&=&\psi^{ }_{-}(t,t)=0.\label{Sol_Sym_End_b}
\end{eqnarray}
Note that for $n\neq 0$, $\psi^{ }_{+}(n,t) \neq \psi^{ }_{-}(n,t)$. In fact $\psi^{ }_{+}(n,t)=\psi^{ }_{-}(n-2,t)$, see Fig.~\ref{Fig_Unbiased}, a property whose implications I discuss below.  Once one has the explicit expression for the components of the wave function, the coin weights read  
\begin{eqnarray}
\cos \theta^{ }_{n,t} &=& \frac{1}{2}\left(\sqrt{1+\frac{n}{t}}-\sqrt{1-\frac{n}{t}}\right),\\
\sin \theta^{ }_{n,t} &=& \frac{1}{2}\left(\sqrt{1+\frac{n}{t}} +\sqrt{1-\frac{n}{t}}\right).
\end{eqnarray}

\section{Biased walk}
\label{Sec_Biased}

\begin{figure}[htbp]
\includegraphics[width=0.9\columnwidth,keepaspectratio=true]{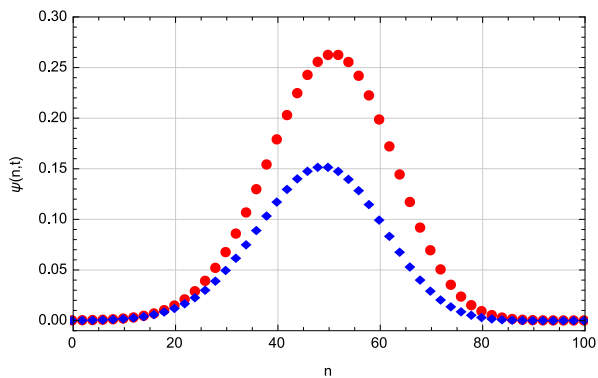}\
\caption{(Color online) 
The two components of the wave function at $t=100$, for $p=0.75$. The red dots denote $\psi^{ }_{+}(n,t)$ whereas the blue diamonds designate the values of $\psi^{ }_{-}(n,t)$.} 
\label{Fig_Biased}
\end{figure}

All these results can be easily modified to encompass the generic case: we simply need to replace the factor $2^{-t}$ in Eqs.~\eqref{Sol_Sym_Psi_P} and~\eqref{Sol_Sym_Psi_M} by the proper combination of powers of $p$ and $(1-p)$. Moreover, conditions~\eqref{Sol_Sym_End_a} and~\eqref{Sol_Sym_End_b} should be mapped into
\begin{eqnarray*}
\psi^{ }_{+}(t,t)&=&p^{\frac{t}{2}},\\
\psi^{ }_{+}(-t,t)&=&\psi^{ }_{-}(t,t)=0,\\
\psi^{ }_{-}(-t,t)&=&\left(1-p\right)^{\frac{t}{2}},
\end{eqnarray*}
what suggests the choice 
\begin{eqnarray}
\psi^{ }_{+}(n,t)&=& \sqrt{\frac{(t-1)!}{\left(\frac{t+n-2}{2}\right)! \left(\frac{t-n}{2}\right)!}}p^{\frac{t+n}{4}}\left(1-p\right)^{\frac{t-n}{4}}, \label{Sol_Gen_Psi_P}\\
\psi^{ }_{-}(n,t)&=&  \sqrt{\frac{(t-1)!}{\left(\frac{t+n}{2}\right)! \left(\frac{t-n-2}{2}\right)!}}p^{\frac{t+n}{4}}\left(1-p\right)^{\frac{t-n}{4}}, \label{Sol_Gen_Psi_M} 
\end{eqnarray}
for $n\in\left\{-t+2,-t+4,\cdots,t-4,t-2\right\}$, see Fig.~\ref{Fig_Biased}. In other words, Eq.~\eqref{Redund_Sym} now splits into 
\begin{eqnarray}
\psi^{ }_{+}(n,t)&=& \sqrt{p} \sqrt{\rho(n-1,t-1)}, \label{Redund_Psi_P}\\
\psi^{ }_{-}(n,t)&=&  \sqrt{1-p} \sqrt{ \rho(n+1,t-1)}. \label{Redund_Psi_M} 
\end{eqnarray}
Finally, we have to use recursive Eqs.~\eqref{Rec_P} and~\eqref{Rec_M} to isolate $\cos \theta^{ }_{n,t}$ and $\sin \theta^{ }_{n,t}$:
\begin{eqnarray}
\cos \theta^{ }_{n,t} &=& \sqrt{\frac{p}{2}}\sqrt{1+\frac{n}{t}}- \sqrt{\frac{1-p}{2}}\sqrt{1-\frac{n}{t}},\label{Cos_Gen}\\
\sin \theta^{ }_{n,t} &=& \sqrt{\frac{1-p}{2}}\sqrt{1+\frac{n}{t}} + \sqrt{\frac{p}{2}}\sqrt{1-\frac{n}{t}},\label{Sim_Gen}
\end{eqnarray}
which satisfy all the desired constraints. Note how expressions~\eqref{Cos_Gen} and~\eqref{Sim_Gen} are ill defined for $n=t=0$: in fact, Eqs.~\eqref{Cos_Sym} and \eqref{Sin_Sym} evidenced this same issue. To be consequent with the previous setup and, in particular, with Eq.~\eqref{Psi_Zero_Gen}, the right option is the most obvious, i.e.,
\begin{eqnarray}
\cos \theta^{ }_{0,0} &=& \sqrt{\frac{p}{2}}- \sqrt{\frac{1-p}{2}},\\
\sin \theta^{ }_{0,0} &=& \sqrt{\frac{1-p}{2}} + \sqrt{\frac{p}{2}},
\end{eqnarray}
but since our coin operator is time dependent, one could modify $\theta^{ }_{0,0}$ and $|\psi\rangle_{0}$ at will, as long as one has
\begin{equation}
|\psi\rangle_{1} =\sqrt{p}\,|+\rangle\otimes | 1\rangle +\sqrt{1-p}\, |-\rangle\otimes | -1\rangle,
\label{psi_one_gen}
\end{equation}
unchanged. This invariance is just one of the many possible transformations that preserves the functional form of $\rho(n,t)$ \cite{MBD14}, but this will be the subject of future research.

\section{Redundancy and coherence}
\label{Sec_Redundancy}

Consider now the following prominent consequence of Eqs.~ \eqref{Redund_Psi_P} and ~\eqref{Redund_Psi_M}. On the one hand, we have a high degree of redundancy, with almost the same information stored in each component of the wave function. On the other hand, this information is the PMF of the system {\it one time step before\/}. All together implies that one can undo the consequences of an accidental measurement of the chirality at time $t$, {\it by means of unitary transformations\/}. In particular, if $|\psi\rangle_{t} \to |\widetilde{\psi}^{+}\rangle_{t}$, one has
\begin{eqnarray}
\widetilde{\psi}^{+}_{+}(n,t)&=& \sqrt{\frac{(t-1)!}{\left(\frac{t+n-2}{2}\right)! \left(\frac{t-n}{2}\right)!}}p^{\frac{t+n-2}{4}}\left(1-p\right)^{\frac{t-n}{4}}, \label{Measured_Psi_P}\\
\widetilde{\psi}^{+}_{-}(n,t)&=& 0, \label{Measured_Psi_M} 
\end{eqnarray}
and the recovery procedure is
\begin{equation}
|\psi\rangle_{t} =\widehat{L}\widehat{S}\widehat{V}^{+}|\widetilde{\psi}^{+}\rangle_{t},
\label{recover_P}
\end{equation}
where $\widehat{V}^{+}$,
\begin{eqnarray}
\widehat{V}^{+}&\equiv&\big[\sqrt{p} |+\rangle  \langle +| +\sqrt{1-p}  |+\rangle  \langle -| \nonumber \\
&+&\sqrt{1-p}   |-\rangle  \langle +| - \sqrt{p}  |-\rangle  \langle -|\big]\otimes\widehat{I}^{\text p},
\end{eqnarray}
is a homogeneous coin operator, and $\widehat{L}$,
\begin{equation}
\widehat{L}\equiv \widehat{I}^{\text c}\otimes\sum_{n=-\infty}^{\infty} |n-1\rangle  \langle n|,
\end{equation}
represents a systematic shift {\it to the left\/}. Thus, the joint operation of $\widehat{L}\widehat{S}$ displaces the negative component of the wave function two sites to the left, whereas the positive component remains in place.
On the contrary, if one has obtained $|\widetilde{\psi}^{-}\rangle_{t}$, the unitary operation is
\begin{equation}
|\psi\rangle_{t} =\widehat{R}\widehat{S}\widehat{V}^{-}|\widetilde{\psi}^{-}\rangle_{t},
\label{recover_M}
\end{equation}
with
\begin{eqnarray}
\widehat{V}^{-}&\equiv&\big[-\sqrt{1-p}|+\rangle  \langle +| + \sqrt{p}  |+\rangle  \langle -| \nonumber \\
&+& \sqrt{p}   |-\rangle  \langle +| +\sqrt{1-p}   |-\rangle  \langle -|\big]\otimes\widehat{I}^{\text p},
\end{eqnarray}
and
\begin{equation}
\widehat{R}\equiv \widehat{I}^{\text c}\otimes\sum_{n=-\infty}^{\infty} |n+1\rangle  \langle n|.
\end{equation}
In the last expressions $\widehat{I}^{\text c}$ and $\widehat{I}^{\text p}$ denoted the identity operator of the corresponding Hilbert space. 

The procedure just described can revert the system to the previous state provided that the outcome of the accidental measurement of the chirality is known. Otherwise, the quantum walker will suffer decoherence since there is some probability that one chooses the erroneous unitary transformation, i.e., that one applies Eq.~\eqref{recover_P} when the right choice is~\eqref{recover_M}, and vice versa. 

Consider, for instance, that the fortuitous measurement occurs after the first time step yielding
\begin{equation*}
|\widetilde{\psi}^{-}\rangle_{1}=|-\rangle\otimes | -1\rangle,
\label{psi_one_measurement}
\end{equation*}
and that we perform the wrong unitary transformation:
\begin{eqnarray}
|\phi\rangle_{1} &\equiv&\widehat{L}\widehat{S}\widehat{V}^{+}|\widetilde{\psi}^{-}\rangle_{1}\nonumber \\
&=&\sqrt{1-p}\,|+\rangle\otimes | -1\rangle -\sqrt{p}\, |-\rangle\otimes | -3\rangle.
\label{phi_one}
\end{eqnarray}
The new PMF is clearly different from the one we had but in a way that is difficult to quantify. Standard methods of information theory, as the Kullback-Leibler divergence~\cite{KL51}, require that the null sets of both probability measures are equal, which is not the case here, cf. Eqs.~\eqref{psi_one_gen} and~\eqref{phi_one}. One cannot resort to simple quantum arguments to establish the resemblance between $ |\phi\rangle_{t}$ and $|\psi\rangle_{t}$ either: Note that as 
\begin{equation*}
\langle \widetilde{\psi}^+ |\widetilde{\psi}^-\rangle_{t}=0, 
\end{equation*}
one must have necessarily
\begin{equation*}
\langle \psi |\phi\rangle_{t}=0,
\end{equation*}
since the recovery procedure is unitary.

The potential relevance of an incidental and unnoticed measurement of the coin state will drastically depend on the actual correlation between the two components of the wave function: the higher the correlation, the lower the consequences. In other words, the impact will decrease with the actual level of entanglement between chiral and spatial degrees of freedom of the original quantum state. 

A way to assess this level of entanglement of the walker is through the entropy of entanglement~\cite{CLXGKK05,ASRD06}, a special instance of von Neumann's entropy. The von Neumann entropy  $S(t)$ of a quantum system is defined in analogy of the Gibbs entropy by:
\begin{equation}
S(t)\equiv-\mbox{tr}\left(\widehat{\rho}_t \log_2 \widehat{\rho}_t\right),
\end{equation} 
where $\widehat{\rho}_t$ is the density matrix operator at time $t$, and $\mbox{tr}\left(\cdot\right)$ is the trace, e.g.,  
\begin{eqnarray}
\mbox{tr}\left(\widehat{\rho}_t\right)=\sum_{n=-\infty}^{\infty}\langle n| \otimes \big[\langle +| \widehat{\rho}_t |+\rangle +\langle -| \widehat{\rho}_t |-\rangle \big] \otimes |n\rangle.\nonumber\\
\end{eqnarray}
In our case, since the time evolution before the accidental measurement is unitary, we will have
\begin{equation}
\widehat{\rho}_t=|\psi\rangle_{t}\langle\psi|,
\end{equation} 
and consequently $S(t)=0$. However, as we are interested in quantifying the entanglement intensity between chirality and position, one can use the {\it reduced\/} von Neumann entropy:  
\begin{equation}
S^{\text c}(t)\equiv-\mbox{tr}^{\text c}\left(\widehat{\rho}^{\text c}_t \log_2 \widehat{\rho}^{\text c}_t\right),
\end{equation} 
where $\widehat{\rho}^{\text c}_t$ is the reduced density matrix operator obtained when a partial trace is taken over the positions:
\begin{eqnarray}
\widehat{\rho}^{\text c}_t
&=& P^{ }_{+}(t) |+\rangle  \langle +| + Q(t) |+\rangle  \langle -| \nonumber \\
&+&Q(t)  |-\rangle  \langle +| + P^{ }_{-}(t) |-\rangle  \langle -|,
\label{reduced_density}
\end{eqnarray}
with~\cite{AR10,AR12}:
\begin{eqnarray}
P^{ }_{+}(t)&\equiv&\sum_{n=-t}^{t} \left| \psi^{ }_{+}(n,t)\right|^2,\\
P^{ }_{-}(t)&\equiv&\sum_{n=-t}^{t} \left| \psi^{ }_{-}(n,t)\right|^2,\\
Q(t)&\equiv&\sum_{n=-t}^{t}  \psi^{ }_{+}(n,t)\psi^{ }_{-}(n,t),
\end{eqnarray}
and $\mbox{tr}^{\text c}(\cdot)$ is the trace restricted to $\HH^{\text c}$. Here
\begin{eqnarray}
P^{ }_{+}(t)&=&p,\\
P^{ }_{-}(t)&=&1-p,
\end{eqnarray}
for all $t$, see Eqs.~\eqref{Redund_Psi_P} and~\eqref{Redund_Psi_M}, and
\begin{eqnarray}
Q(t)&=&\sqrt{p(1-p)}\sum_{n=-t}^{t} \sqrt{\rho(n-1,t-1)\rho(n+1,t-1)}. \nonumber\\
\end{eqnarray}

Under these circumstances, the entropy of entanglement can be expressed in terms of $\lambda_{\pm}^{\text c}(t)$, the eigenvalues of the reduced density matrix at time $t$,
\begin{equation}
S^{\text c}(t)=-\lambda_+^{\text c}(t) \log_2 \lambda_+^{\text c}(t)-\lambda_-^{\text c}(t) \log_2 \lambda_-^{\text c}(t),
\end{equation} 
with,
\begin{eqnarray}
\lambda_{\pm}^{\text c}(t)=\frac{1}{2}\pm\sqrt{\frac{1}{4}-p(1-p)+Q^2(t)}.
\end{eqnarray}

\begin{figure}[htbp]
\includegraphics[width=0.9\columnwidth,keepaspectratio=true]{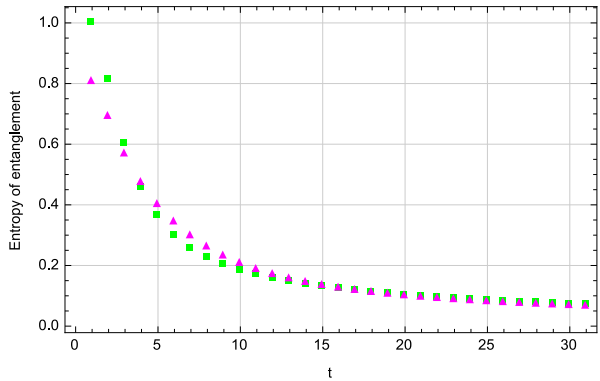}
\caption{(Color online) 
Evolution of the entropy of entanglement as a function of time for $p=0.5$, green boxes, and $p=0.75$, magenta triangles. 
Observe how the entropy decreases monotonically in both cases.} 
\label{Fig_Entropy}
\end{figure}

Figure~\ref{Fig_Entropy} shows the values of the entropy for $t\geq1$, for the two examples considered in previous sections, the unbiased walk, $p=0.5$, and the biased one, $p=0.75$. In both instances, the entanglement is maximal for $t=1$, when there are no off-diagonal terms in $\widehat{\rho}^{\text c}_t$, and one has a one-to-one equivalence between the information carried by  chirality and the spatial location |this just the case of the explicit example shown above, Eqs.~\eqref{psi_one_gen} and~\eqref{phi_one}. After that point, 
the magnitude of the entanglement in both cases converges and decreases monotonically toward zero: there is less information susceptible of getting lost.      

This conclusion can also be derived from the direct analysis of $Q(t)$ for $t\gg1$.  In this regime, one can approximate
\begin{equation*}
\rho(n-1,t-1)\sim\rho(n+1,t-1), 
\end{equation*}
so that
\begin{eqnarray}
\lim_{t\to\infty} Q(t)=\sqrt{p(1-p)},
\end{eqnarray}
and thus
\begin{eqnarray}
\lim_{t\to\infty} S^{\text c}(t)=0.
\end{eqnarray}

\begin{figure}[htbp]
\includegraphics[width=0.9\columnwidth,keepaspectratio=true]{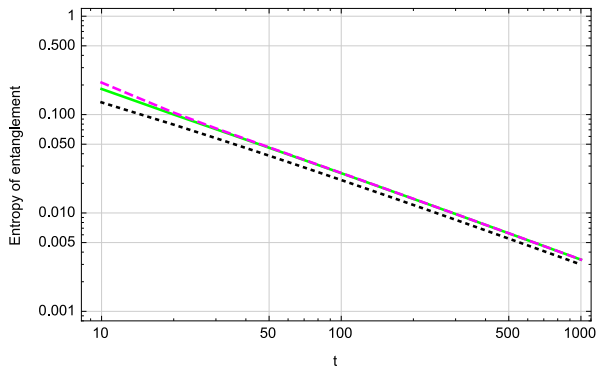}
\caption{(Color online) 
Decay of the entropy of entanglement as a function of time. 
The green solid line corresponds to $p=0.5$, whereas the magenta dashed line shows the behavior of the entanglement when $p=0.75$. The black dotted line ($\sim [\log_2 4t]/4 t$) serves as a guide for the eye.} 
\label{Fig_Decay}
\end{figure}

A more detailed analysis reveals that the leading term of the reduced entropy is of the form
\begin{eqnarray}
S^{\text c}(t)\sim \frac{1}{4 t}\log_2 4t,
\end{eqnarray}
and therefore it does not depend on the value of $p$, as it can be seen in Fig.~\ref{Fig_Decay}.

\section{Decoherence}
\label{Sec_Decoherence}

\begin{figure}[htbp]
\includegraphics[width=0.9\columnwidth,keepaspectratio=true]{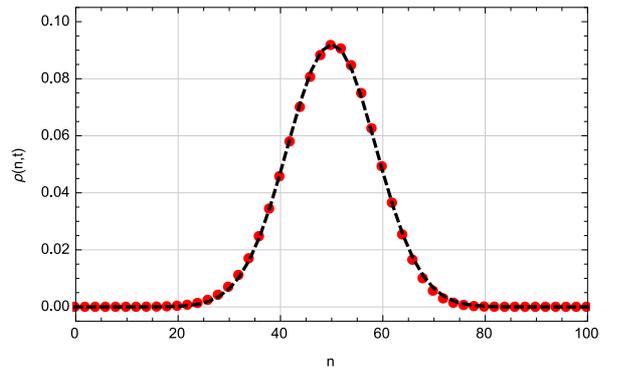}
\caption{(Color online) 
Probability mass function $\rho(n,t)$ at $t=100$, for $p=0.75$. Red dots were obtained from the wave function of the coherent process whereas the black dashed line corresponds to a Gaussian approximation, Eq.~\eqref{Gauss_approx}.}
\label{Fig_Gauss}
\end{figure}

From the analysis above, one can conclude that the present QW is particularly resistant to decoherence, but by no means it is immune to it. This is a somewhat paradoxical situation since the general rule dictates that quantum walks turn into random walks when decohered~\cite{VK07}, and our starting point is a process whose distribution already coincides with the classical one: when $t$ is large enough 
\begin{equation}
\rho(n,t)\to \frac{2}{\sqrt{2 \pi \sigma^2 t}}e^{-\frac{(n-\mu t)^2}{2 \sigma^2 t}},
\label{Gauss_approx}
\end{equation}
where $\mu=2p-1$, $\sigma^2=4p(1-p)$, and the factor $2$ is due to the odd/even alternating nature of $n$ |see Fig~\ref{Fig_Gauss}. 

For homogeneous, time-independent coins as, e.g., the Hadamard coin, the introduction of random measurements on both position and chirality of the quantum particle is enough to induce the shift from the original ballistic behavior of the system, with a variance that scales as $t^2$, to the classical behavior of a random walk with a variance that grows linearly with time, and whose exact expression is a function of the measurement probability per unit of time, $q$~\cite{KZ08}. This approach does not suffice in the present case, due to the aforementioned existing link between chiral and positional degrees of freedom. 

Hence, we have to remove any residual correlation in the system to ensure the validity of the central limit theorem: We will perform measurements of the location of the particle at random times and, after having pinned down its position, we will restore the chirality of the system to its initial state~\cite{SCSK10}. With this method, our QW turns into a sum of independent random variables which are {\it not\/} identically distributed, since the coin operator is still site- and time-dependent.~\footnote{In some practical implementations of the discrete-time QW, the role of the time variable is assumed by an auxiliary spatial dimension, what breaks in practice temporal homogeneity~\cite{SWH14}.} 

This lack of homogeneity in the coin operator implies that the random walk obtained through decoherence will be a {\it biased\/} random walk, even if $p=0.5$. This bias may lead to limiting probability density functions that are Gaussian, but it is well known~\cite{MM07} that depending on the strength of persistence, bimodal distributions may appear: Hence, in general, we will not recover the same statistical properties of the process whose evolution we were mimicking. 

\begin{figure}[htbp]
\includegraphics[width=0.9\columnwidth,keepaspectratio=true]{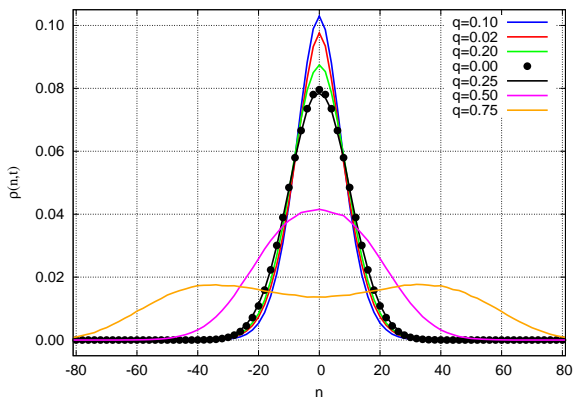}
\caption{(Color online) 
Probability mass function $\rho(n,t)$ for $t=100$, $p=0.5$, and different values of the measurement probability $q$. Data were obtained by averaging 1,000,000 different realizations of the process, except for the case $q=0.00$ that corresponds to the coherent walk. Labels are ordered to indicate increasing variance.}
\label{Fig_Decoherent}
\end{figure}

We can observe all these phenomena in Fig.~\ref{Fig_Decoherent}. Small measurement probabilities, $q\ll 0.1$, result in normal distributions with variance reduction. After that point approximately, $q\gtrsim 0.1$, variance starts increasing with $q$, and eventually becomes larger than the original value. This happens for measurement probabilities $q\gtrsim 0.25$ |see the collapse of the plots corresponding to $q=0$ and $q=0.25$ in Fig.~\ref{Fig_Decoherent}. Finally, for values of the measurement probability of about $q\gtrsim 0.7$, bimodal distributions do appear.

\section{Conclusion}
\label{Sec_Conclusion}

Inspired by the fact that quantum walks are universal computation primitives, and thus they can solve any problem that can be tackled by a general-purpose computer, I looked for a particular instance that reproduced the statistical features of a random walk. 

The aim was to design a non-trivial version of the discrete-time quantum walk on the line with exactly the same probability of site occupation as the classical process at any time scale, not as a byproduct of the asymptotic loss of coherence in the quantum evolution. Along the text, I have proved that one possible way to get the desired behavior is through the introduction of an inhomogeneous, time-dependent coin operator.   

The correlation level shown by both components of the wave function is so high that one can use it to restore the system to the same state previous to a measurement of its chirality. This perfect reversion can be performed with the only aid of unitary operators whenever one knows the output of the measuring process. Moreover, the analysis of the entropy of entanglement between positional and chiral degrees of freedom shows that the latter information loses significance as time increases. 

Finally, this restoring procedure can be seen as a simple protection mechanism against accidental degradation of the coherence of the quantum state, but it can lead to some other yet undiscovered interesting implications.

\acknowledgments
The author acknowledges support from the Spanish MINECO 
under Contract No. FIS2013-47532-C3-2-P, and from AGAUR, 
Contract No. 2014SGR608.



\begin{thebibliography}{00}


\bibitem{VA12} \Journal{S. E. Venegas-Andraca}{2012}{Quantum walks: a comprehensive review}{Quantum Inf. Process.}{11}{1015}{106} 

\bibitem{ADZ93} \Journal{Y. Aharonov, L. Davidovich, and N. Zagury}{1993}{Quantum random walks}{Phys. Rev. {\rm A}}{48}{1687}{90} 

\bibitem{TM02} \JournalE{B. C. Travaglione and G. J. Milburn}{2002}{Implementing the quantum random walk}{Phys. Rev. {\rm A}}{65}{032310} 

\bibitem{NK03} \Journal{N. Konno}{2003}{Quantum Random Walks in One Dimension}{Quantum Inf. Process.}{1}{345}{54} 

\bibitem{JK03} \Journal{J. Kempe}{2003}{Quantum random walks: An introductory overview}{Contemp. Phys.}{44}{307}{27} 


\bibitem{ABNVW01} \Bookwd{A. Ambainis, E. Bach, A. Nayak, A. Vishwanath, and J. Watrous}{2001}{{\rm in} One Dimensional Quantum Walks, {\rm Proceedings of the thirty-third annual ACM symposium on Theory of Computing}}{ACM New York}{New York}, p. 37. 

\bibitem{CFG03} \Journal{A. Childs, E. Farhi, and S. Gutmann}{2003}{An Example of the Difference Between Quantum and Classical Random Walks}{Quantum Inf. Process.}{1}{35}{43} 


\bibitem{BCA03a} \JournalE{T. A. Brun, H. A. Carteret,  and A. Ambianis}{2003}{Quantum random walks with decoherent coins}{Phys. Rev. {\rm A}}{67}{032304} 

\bibitem{BCA03b} \JournalE{T. A. Brun, H. A. Carteret, and A. Ambainis}{2003}{Quantum to Classical Transition for Random Walks}{Phys. Rev. Lett.}{91}{3130602} 

\bibitem{KT03} \JournalE{V. Kendon and B. Tregenna}{2003}{Decoherence can be useful in quantum walks}{Phys. Rev. {\rm A}}{67}{042315} 

\bibitem{WD03} \JournalE{A. W\'ojcik and J. R. Dorfman}{2003}{Diffusive-Ballistic Crossover in 1D Quantum Walks}{Phys. Rev. Lett.}{90}{230602} 

\bibitem{SCPGJS11} \JournalE{A. Schreiber, K. N. Cassemiro, V. Poto\v{c}ek, A. G\'abris, I. Jex, and Ch. Silberhorn}{2011}{Decoherence and Disorder in Quantum Walks: From Ballistic Spread to Localization}{Phys. Rev. Lett.}{106}{180403} 


\bibitem{AVWW11} \JournalE{A. Ahlbrecht, H. Vogts, A. H. Werner, and R. F. Werner}{2011}{Asymptotic evolution of quantum walks with random coin}{J. Math. Phys.}{52}{042201} 

\bibitem{ACMSWW2} \Journal{A. Ahlbrecht, C. Cedzich, V. B. Scholz, A. H. Werner, and R. F. Werner}{2012}{Asymptotic behavior of quantum walks with spatio-temporal coin fluctuations}{Quantum Inf. Process.}{11}{1219}{49} 



\bibitem{RPF85} \Journal{R. P. Feynman}{1985}{Quantum Mechanical Computers}{Opt. News}{11}{11}{20}  

\bibitem{DAM96} \Journal{D. A. Meyer}{1996}{From Quantum Cellular Automata to Quantum Lattice Gases}{J. Stat. Phys}{85}{551}{74}  


\bibitem{PS97} \Journal{P. W. Shor}{1997}{Polynomial-time algorithms for prime factorization and discrete logarithms on a quantum computer}{SIAM J. Comp.}{26}{1484}{509} 

\bibitem{LKG97} \Journal{L. K. Grover}{1997}{Quantum Mechanics Helps in Searching for a Needle in a Haystack}{Phys. Rev. Lett.}{79}{325}{28} 

\bibitem{FG98} \Journal{E. Farhi and S. Gutmann}{1998}{Quantum computation and decision trees}{Phys. Rev. {\rm A}}{58}{915}{28} 


\bibitem{SKW03} \JournalE{N. Shenvi, J. Kempe, and K. B. Whaley}{2003}{Quantum random-walk search algorithm}{Phys. Rev. {\rm A}}{67}{052307} 

\bibitem{AMB10} \JournalE{E. Agliari, A. Blumen, and O. N\"ulken}{2010}{Quantum-walk approach to searching on fractal structures}{Phys. Rev. {\rm A}}{82}{012305} 

\bibitem{MNRS11} \Journal{F. Magniez, A. Nayak, J. Roland, and M. Santha}{2011}{Search via quantum walk}{SIAM J. Comp.}{40}{142}{64} 


\bibitem{AC09} \JournalE{A. M. Childs}{2009}{Universal Computation by Quantum Walk}{Phys. Rev. Lett.}{102}{180501} 

\bibitem{LCETK10} \JournalE{N. B. Lovett, S. Cooper, M. Everitt, M. Trevers, and V. Kendon}{2010}{Universal quantum computation using the discrete-time quantum walk}{Phys. Rev. {\rm A}}{81}{042330} 

\bibitem{DD85} \Journal{D. Deutsch}{1985}{Quantum Theory, the Church-Turing Principle and the Universal Quantum Computer}{Proc. R. Soc. A-Math. Phys. Eng. Sci.}{400}{97}{117} 

\bibitem{NC10} \Book{M. A. Nielsen, and I. L. Chuang}{2010}{Quantum Computation and Quantum Information: 10th Anniversary Edition}{Cambridge University Press}{New York}


\bibitem{BFT08} \JournalE{D. Bulger, J. Freckleton, and J. Twamley}{2008}{Position-dependent and cooperative quantum Parrondo walks}{New J. Phys.}{10}{093014} 

\bibitem{SK10} \JournalE{Y. Shikano and H. Katsura}{2010}{Localization and fractality in inhomogeneous quantum walks with self-duality}{Phys. Rev. {\rm E}}{82}{031122} 

\bibitem{KLS13} \Journal{N. Konno, T. \L uczak, and E. Segawa}{2013}{Limit measures of inhomogeneous discrete-time quantum walks in one dimension}{Quantum Inf. Process.}{12}{33}{53} 

\bibitem{ZXT14} \JournalE{R. Zhang, P. Xue, and J. Twamley}{2014}{One-dimensional quantum walks with single-point phase defects}{Phys. Rev. {\rm A}}{89}{042317} 

\bibitem{XQTS14} \JournalE{P. Xue, H. Qin, B. Tang, and B. C. Sanders}{2014}{Observation of quasiperiodic dynamics in a one-dimensional quantum walk of single photons in space}{New J. Phys.}{16}{053009} 


\bibitem{RMM04} \JournalE{P. Ribeiro, P. Milman, and R. Mosseri}{2004}{Aperiodic Quantum Random Walks}{Phys. Rev. Lett.}{93}{190503} 

\bibitem{BNPRS06} \JournalE{M. C. Ba\~nuls, C. Navarrete, A. P\'erez, E. Rold\'an, and J. C. Soriano}{2006}{Quantum walk with a time-dependent coin}{Phys. Rev. {\rm A}}{73}{062304} 

\bibitem{AR09a} \Journal{A. Romanelli}{2009}{The Fibonacci quantum walk and its classical trace map}{Physica {\rm A}}{388}{3985}{90}

\bibitem{AR09b} \JournalE{A. Romanelli}{2009}{Driving quantum-walk spreading with the coin operator}{Phys. Rev. {\rm A}}{80}{042332} 

\bibitem{MM14} \JournalE{M. Montero}{2014}{Invariance in quantum walks with time-dependent coin operators}{Phys. Rev. {\rm A}}{90}{062312} 


\bibitem{FAJ04} \Journal{A. P. Flitney, D. Abbott, and N. F. Johnson}{2004}{Quantum walks with history dependence}{J. Phys. {\rm A}}{37}{7581}{91} 

\bibitem{SWH14} \JournalE{Y. Shikano, T. Wada, and J. Horikawa}{2014}{Discrete-time quantum walk with feed-forward quantum coin}{Sci. Rep.}{4}{4427} 


\bibitem{TFMK03} \JournalE{B. Tregenna, W. Flanagan, R. Maile, and V. Kendon}{2003}{Controlling discrete quantum walks: coins and initial states}{New J. Phys.}{5}{83} 

\bibitem{MM15} \Journal{M. Montero}{2015}{Quantum walk with a general coin: exact solution and asymptotic properties}{Quantum Inf. Process.}{14}{839}{66} 

\bibitem{MBD14} \Journal{G. Di Molfetta, M. Brachet, and F. Debbasch}{2014}{Quantum walks in artificial electric and gravitational fields}{Physica {\rm A}}{397}{157}{168} 


\bibitem{CLXGKK05} \JournalE{I. Carneiro, M. Loo, X. Xu, M. Girerd, V. Kendon, and P. L. Knight}{2005}{Entanglement in coined quantum walks on regular graphs}{New J. Phys.}{7}{156}

\bibitem{ASRD06} \JournalE{G. Abal, R. Siri, A. Romanelli, and R. Donangelo}{2006}{Quantum walk on the line: Entanglement and nonlocal initial conditions}{Phys. Rev. {\rm A}}{73}{042302}  

\bibitem{KL51} \Journal{S. Kullback, and R. A. Leibler}{1951}{On Information and Sufficiency}{Ann. Math. Statist.}{22}{79}{86} 

\bibitem{AR10} \JournalE{A. Romanelli}{2010}{Distribution of chirality in the quantum walk: Markov process and entanglement}{Phys. Rev. {\rm A}}{81}{062349} 

\bibitem{AR12} \JournalE{A. Romanelli}{2012}{Thermodynamic behavior of the quantum walk}{Phys. Rev. {\rm A}}{85}{012319} 


\bibitem{VK07} \Journal{V. Kendon}{2007}{Decoherence in Quantum Walks - a review}{Math. Struct. Comp. Sci.}{17}{1169}{220} 

\bibitem{KZ08} \JournalE{K. Zhang}{2008}{Limiting distribution of decoherent quantum random walks}{Phys. Rev. {\rm A}}{77}{062302} 

\bibitem{SCSK10} \JournalE{Y. Shikano, K. Chisaki, E. Segawa, and N. Konno}{2010}{Emergence of randomness and arrow of time in quantum walks}{Phys. Rev. {\rm A}}{81}{062129} 


\bibitem{MM07} \JournalE{M. Montero and J. Masoliver}{2007}{Nonindependent continuous-time random walks}{Phys. Rev. {\rm E}}{76}{061115} 

\end{thebibliography}
\end{document}